\documentclass[pdflatex,sn-mathphys]{sn-jnl}

\usepackage{bm}
\DeclareMathOperator{\Tr}{Tr}

\jyear{2021}%

\theoremstyle{thmstyleone}%
\theoremstyle{thmstyletwo}%

\theoremstyle{thmstylethree}%

\begin{document}

\title[Violation of Ericksen inequalities in lyotropic chromonic liquid crystals]{Violation of Ericksen inequalities in lyotropic chromonic liquid crystals}


\author[1]{\fnm{Cheng} \sur{Long}}\nomail

\author*[1]{\fnm{Jonathan V.} \sur{Selinger}}\email{jselinge@kent.edu}

\affil[1]{\orgdiv{Department of Physics, Advanced Materials and Liquid Crystal Institute}, \orgname{Kent State University}, \orgaddress{\city{Kent}, \state{Ohio}, \postcode{44242}, \country{USA}}}


\abstract{By analyzing elastic theory for nematic liquid crystals, we distinguish three regimes of elastic constants.  In one regime, the Ericksen inequalities are satisfied, and the ground state of the director field is uniform.  In a second regime, certain necessary inequalities are violated, and the free energy is thermodynamically unstable.  Between those possibilities, there is an intermediate regime, where the Ericksen inequalities are violated but the system is still stable.  Remarkably, lyotropic chromonic liquid crystals are in the intermediate regime.  We investigate the nonuniform structure of the director field in that regime, show that it depends sensitively on system geometry, and discuss the implications for lyotropic chromonic liquid crystals.}

\maketitle

\section{Introduction}

In a nematic liquid crystal, the molecules have orientational order along an axis, which is called the nematic director and is represented by the unit vector $\hat{\bm{n}}$.  The molecules are equally likely to orient up or down along this axis, and hence $\hat{\bm{n}}$ and $-\hat{\bm{n}}$ represent the same physical state.  The axis of orientational order normally depends on position, so that the director becomes a position-dependent field $\hat{\bm{n}}(\bm{r})$.  One fundamental problem of liquid-crystal elasticity theory is to determine how the free energy depends on spatial gradients of the director field.  In early studies of liquid crystals, this problem was addressed by Oseen~\cite{Oseen1933} and Frank~\cite{Frank1958}, and further by Nehring and Saupe~\cite{Nehring1971,Nehring1972}.  Their work led to the Oseen-Frank free energy density, which includes terms representing three deformation modes---splay, twist, and bend---as well as a fourth term called saddle-splay.  Hence, it has four independent elastic coefficients.

In any elasticity theory, there must be certain inequalities on the elastic coefficients so that the free energy is thermodynamically stable; i.e.\ no deformation can make the free energy go to negative infinity.  For example, in conventional elasticity for an isotropic solid, the bulk and shear moduli must both be positive.  Soon after the development of the Oseen-Frank free energy density for liquid crystals, Ericksen~\cite{Ericksen1966} derived a corresponding set of four inequalities on the liquid-crystal elastic coefficients.  These four inequalities will be presented in detail below.

The purpose of this paper is to re-examine the Ericksen inequalities.  There are both theoretical and experimental reasons for this re-examination.

The theoretical reason is that our group has recently developed a reformulation of liquid-crystal elasticity theory~\cite{Selinger2018}, which is based on a mathematical decomposition of the director gradient tensor by Machon and Alexander~\cite{Machon2016}.  This reformulation expresses the free energy density in terms of four director deformation modes---splay, twist, bend, and a less-well-known fourth mode represented as the tensor $\bm{\Delta}$.  It is mathematically equivalent to the Oseen-Frank free energy density, but it provides a conceptually simpler way to describe many elastic phenomena in liquid crystals.  Using this reformulation, several groups have investigated geometric compatibility constraints on director deformations~\cite{Virga2019,Sadoc2020,Pollard2021,DaSilva2021}.  As we will show, this theoretical progress provides new insight into the Ericksen inequalities.

The experimental reason is that a violation of the Ericksen inequalities has actually been observed in certain liquid-crystal materials.  These materials are lyotropic chromonic liquid crystals, such as Sunset Yellow (SSY) and disodium cromoglycate (DSCG).  In these materials, the molecules self-organize into long stacks in aqueous solution, and the stacks form nematic orientational order.  Several experiments have put these liquid crystals into cylindrical capillaries~\cite{Nayani2015,Davidson2015}, rectangular capillaries~\cite{Fu2017}, or cylindrical shells~\cite{Javadi2018}, with degenerate planar anchoring on the surfaces.  In these geometries, the director field spontaneously forms a twisted structure rather than a uniform, achiral state.  The twisted structure has been modeled by the experimental groups using a twist elastic constant that is anomalously small, violating one of the Ericksen inequalities.  Based on this observation, one must ask:  Is the violation consistent with liquid-crystal elasticity theory?  If so, are there any constraints on the elastic constants?

In Sec.~2 of this article, we present the four Ericksen inequalities, using both the standard form of the Oseen-Frank free energy and our recent reformulation.  We point out that these inequalities are excessively strict, because of the geometric constraints on the director deformation modes.  In the following sections, we investigate what happens if the Ericksen inequality on the twist elastic constant is violated.  In Sec.~3, we consider a severe violation ($K_{22}<0$), and show that the free energy can become arbitrary negative.  This violation is forbidden for reasons of thermodynamic stability (unless the free energy includes higher-order terms).  In Sec.~4, we consider the intermediate regime ($0<K_{22}<K_{24}$), as in the experiments on lyotropic chromonic liquid crystals, and show that the free energy does not become arbitrary negative.  Rather, the liquid crystal has a well-defined, non-uniform ground state, which depends on the system size.  Finally, in Sec.~5, we discuss the implications of these results for studies of lyotropic chromonic liquid crystals.

\section{Theoretical background}

Oseen-Frank theory shows all the ways that the free energy density can depend on spatial gradients of the director field, as allowed by symmetry, up to quadratic order in the gradients.  The free energy density is conventionally written as
\begin{equation}
    F=\frac{1}{2}K_{11}S^2+\frac{1}{2}K_{22}T^2+\frac{1}{2}K_{33}\lvert\bm{B}\rvert^2
    -K_{24}\bm{\nabla}\cdot\left[\hat{\bm{n}}(\bm{\nabla}\cdot\hat{\bm{n}})
    +\hat{\bm{n}}\times(\bm{\nabla}\times\hat{\bm{n}})\right].
    \label{oseenfrank}
\end{equation}
Here, $S=\bm{\nabla}\cdot\hat{\bm{n}}$ is the splay, $T=\hat{\bm{n}}\cdot(\bm{\nabla}\times\hat{\bm{n}})$ is the twist, and $\bm{B}=-(\hat{\bm{n}}\cdot\bm{\nabla})\hat{\bm{n}}=\hat{\bm{n}}\times(\bm{\nabla}\times\hat{\bm{n}})$ is the bend.  These three deformation modes contribute to the free energy density with elastic constants $K_{11}$, $K_{22}$, and $K_{33}$, respectively.  The fourth term is called the saddle-splay term.  This expression for saddle-splay has been used, for example, by Burylov~\cite{Burylov1997} to model transitions among different director configurations in a cylindrical capillary.

In the literature, there are some variations in the notation for the saddle-splay coefficient:  it sometimes written as $K_{24}$,  $\frac{1}{2}K_{24}$, $({K_{22}+K_{24}})$, or $\frac{1}{2}({K_{22}+K_{24}})$.  We choose the notation of Eq.~(\ref{oseenfrank}), because it seems to be the most common in recent articles, but our results can easily be translated into any of the other notations.

In Ref.~\cite{Selinger2018}, we argue that the elastic free energy for nematic liquid crystals can more easily be understood in terms of four bulk elastic modes, rather than three.  This argument is based on a mathematical construction of Machon and Alexander~\cite{Machon2016}, who decompose the director gradient tensor into four different types of mathematical objects as
\begin{equation}
    \partial_i n_j = -n_i B_j + \frac{1}{2}S(\delta_{ij}-n_i n_j) + \frac{1}{2}T\epsilon_{ijk}n_k + \Delta_{ij}.
\end{equation}
Here, the scalar $S$, pseudoscalar $T$, and vector $\bm{B}$ are the splay, twist, and bend modes defined above.  The fourth mode $\bm{\Delta}$ is a symmetric, traceless tensor in the plane perpendicular to $\hat{\bm{n}}$.  It indicates how the director splays outward in one direction and inward in the orthogonal direction in that plane.  All four modes are visualized in Ref.~\cite{Selinger2018}.  In this theoretical approach, pure splay is double splay, and pure twist is double twist.  Planar single splay is a combination of pure splay and $\bm{\Delta}$ mode, while cholesteric single twist is a combination of pure twist and $\bm{\Delta}$ mode.

Using the four bulk elastic modes, the free energy can be written as
\begin{equation}
    F=\frac{1}{2}(K_{11}-K_{24})S^2+\frac{1}{2}(K_{22}-K_{24})T^2+\frac{1}{2}K_{33}\lvert\bm{B}\rvert^2
    +K_{24}\Tr(\bm{\Delta}^2).
    \label{newfreeenergy}
\end{equation}
We emphasize that Eq.~(\ref{newfreeenergy}) is mathematically equivalent to Eq.~(\ref{oseenfrank}), so these expressions for the free energy can be used interchangeably.  However, Eq.~(\ref{newfreeenergy}) has the special feature that it is a sum of squares of elastic modes.  In the sum of squares, the elastic constant for splay is $(K_{11}-K_{24})$, the elastic constant for twist is $(K_{22}-K_{24})$, the elastic constant for bend is $K_{33}$, and the elastic constant for the $\bm{\Delta}$ mode is $2K_{24}$.  In this perspective, those four elastic constants are the fundamental parameters of the theory, while $K_{11}$ and $K_{22}$ are less fundamental: $K_{11}=(K_{11}-K_{24})+K_{24}$ applies to a combination of splay and $\bm{\Delta}$, and $K_{22}=(K_{22}-K_{24})+K_{24}$ applies to a combination of twist and $\bm{\Delta}$.

Ericksen~\cite{Ericksen1966} looked for conditions on the elastic constants such that the free energy would be positive definite.  Translated into our current notation, his inequalities were
\begin{equation}
    K_{11}>\lvert 2K_{24}-K_{11}\rvert,\quad
    K_{22}>\lvert 2K_{24}-K_{22}\rvert,\quad
    K_{33}>0.
\end{equation}
These inequalities can be rewritten as
\begin{equation}
    K_{11}>K_{24},\quad
    K_{22}>K_{24},\quad
    K_{33}>0,\quad
    K_{24}>0.
    \label{ericksen}
\end{equation}
Because Ericksen worked with the Oseen-Frank free energy in a form similar to Eq.~(\ref{oseenfrank}), the inequalities were not obvious; they required a significant derivation.  However, with the reformulated free energy of Eq.~(\ref{newfreeenergy}), the inequalities are almost trivial.  They simply state that the free energy is a sum of squares with \emph{positive} coefficients.

Clearly the Ericksen inequalities are \emph{sufficient} for the free energy to be thermodynamically stable.  If the inequalities are satisfied, then any director deformation has a positive free energy.  The ground state is a uniform director field, with $F=0$.  However, it is not clear whether the Ericksen inequalities are \emph{necessary} for the free energy to be thermodynamically stable.

The main subtlety here is that the four director deformation modes $S$, $T$, $\bm{B}$, and $\bm{\Delta}$ are not independent of each other.  Rather, they must all be derived from the same director field $\hat{\bm{n}}(\bm{r})$.  Only certain combinations of $S$, $T$, $\bm{B}$, and $\bm{\Delta}$ can be derived from the same director field, and these combinations are called compatible, while other combinations are incompatible.

One way to see this issue is by counting degrees of freedom:  $S$ has one degree of freedom (as a scalar), $T$ has one (as a pseudoscalar), $\bm{B}$ has two (as a vector perpendicular to $\hat{\bm{n}}$), and $\bm{\Delta}$ has two (as a traceless, symmetric tensor in the plane perpendicular to $\hat{\bm{n}}$), for a total of six degrees of freedom in the deformation modes.  By comparison, the director field has only two degrees of freedom (as a unit vector).  Hence, we would expect four constraints among the director deformation modes.

An analogous issue occurs in the elastic theory of solids.  The strain tensor is derived from the displacement vector field, but the strain tensor has more degrees of freedom than the displacement vector field.  Hence, there must be compatibility constraints on the strain tensor.  Here, the director field of a liquid crystal is analogous to the displacement field of a solid, and the deformation modes $S$, $T$, $\bm{B}$, and $\bm{\Delta}$ are analogous to the strain tensor of a solid.

In recent years, several groups have investigated how the compatibility constraints for the director deformation modes are related to the geometry of space.  Virga~\cite{Virga2019} found there are only two ways to fill three-dimensional (3D) Euclidean space with constant director deformations:  twist and $\bm{\Delta}$, or bend, twist, and $\bm{\Delta}$.  Sadoc et al.~\cite{Sadoc2020} found that 3D non-Euclidean curved space can be filled with a single director deformation mode, with the right correspondence between the type of curvature and the type of deformation.  Pollard and Alexander~\cite{Pollard2021} and da Silva and Efrati~\cite{DaSilva2021} developed general compatibility conditions for an arbitrary combination of deformation modes and an arbitrary Euclidean or non-Euclidean geometry.

The compatibility conditions derived in Refs.~\cite{Pollard2021,DaSilva2021} are mathematically complex, and we will not attempt to use them directly.  Instead, we will construct director fields in 3D Euclidean geometry, and determine what happens to these director fields if the Ericksen inequalities are violated.

\section{Necessary inequalities}

For a first step, consider a director field with the cholesteric helical structure
\begin{equation}
    \hat{\bm{n}}=(\cos q z,\sin q z,0).
\end{equation}
As shown in Ref.~\cite{Selinger2018}, this structure does not have pure twist; rather, it is a combination of the twist and $\bm{\Delta}$ deformation modes.  It is one of the allowed constant combinations found by Virga~\cite{Virga2019}.  Inserting this director field into the free energy density of Eq.~(\ref{oseenfrank}) or~(\ref{newfreeenergy}) gives $F=\frac{1}{2}K_{22}q^2$.  It involves $K_{22}$, which is a combination of the 
elastic constant $(K_{22}-K_{24})$ for pure twist and the elastic constant $K_{24}$ for the $\bm{\Delta}$ mode.  Hypothetically, suppose that $K_{22}<0$.  In that case, minimizing the free energy would drive $q\to\pm\infty$, and hence $F\to-\infty$.  That situation would be thermodynamically unstable.  Hence, a \emph{necessary} inequality is $K_{22}>0$.  Note that this inequality is \emph{weaker} than the Ericksen inequality on $K_{22}$.

Next, consider a director field with the twist-bend heliconical structure
\begin{equation}
    \hat{\bm{n}}=(\sin\beta\cos q z,\sin\beta\sin q z,\cos\beta).
\end{equation}
This structure is a combination of the bend, twist, and $\bm{\Delta}$ deformation modes; it is the second allowed constant combination.  Inserting this director field into Eq.~(\ref{oseenfrank}) or~(\ref{newfreeenergy}) gives the free energy density 
\begin{equation}
    F=\frac{1}{2}\sin^2\beta(K_{22}\sin^2\beta+K_{33}\cos^2\beta)q^2.
\end{equation}
Hypothetically, suppose that $K_{33}<0$.  In that case, we can choose a cone angle $\beta$ such that the coefficient of $q^2$ is negative.  Minimizing the free energy then drives $q\to\pm\infty$, and hence $F\to-\infty$, which is thermodynamically unstable.  Hence, another necessary inequality is $K_{33}>0$.  That inequality is the same as one of the Ericksen inequalities.

When we construct director fields, we are not limited to constant combinations of deformation modes; we can also investigate non-uniform combinations.  For example, consider the director wave
\begin{equation}
    \hat{\bm{n}}=(\sin\theta(\bm{r}),0,\cos\theta(\bm{r})),\text{ with }
    \theta(\bm{r})=\theta_0\cos(\bm{q}\cdot\bm{r}),
\end{equation}
for small $\theta_0$.  By putting that wave into the free energy density of Eq.~(\ref{oseenfrank}) or~(\ref{newfreeenergy}) and integrating over $\bm{r}$, we obtain the average free energy density
\begin{equation}
    F_\text{average}=\frac{1}{4}(K_{11}q_x^2+K_{22}q_y^2+K_{33}q_z^2)\theta_0^2 + O(\theta_0^4).
\end{equation}
If any of the three coefficients $K_{11}$, $K_{22}$, or $K_{33}$ were negative, then minimization of the free energy would drive the corresponding component of $\bm{q}$ to $\pm\infty$, and hence drive $F\to-\infty$, which would be thermodynamically unstable.  Hence, all three of those coefficients must be positive, and the set of necessary inequalities becomes
\begin{equation}
    K_{11}>0, \quad K_{22}>0, \quad K_{33}>0.
    \label{necessary}
\end{equation}

At this point, some readers might object that there is actually an extensive literature on liquid crystals with $K_{33}<0$, beginning with the work of Dozov~\cite{Dozov2001}.  However, that situation is different because Dozov uses a free energy with higher-order terms, involving either second derivatives or higher powers of first derivatives of the director field.  These higher-order terms prevent the thermodynamic instability, and give a twist-bend nematic phase with finite values of wavevector $q$ and cone angle $\theta$.  Likewise, in recent work by our group~\cite{Shamid2013,Rosseto2020,Selinger2022,Rosseto2022}, an effective renormalized elastic constant ($K_{33}^R$, $K_{11}^R$, or $K_{22}^R$) is driven negative by interactions with another order parameter, and the free energy would be unstable, but it is stabilized by some higher-order couplings.

In elasticity theory, it is always possible to add higher-order terms to stabilize a free energy.  When we search for stability conditions, the issue is whether the free energy is stable \emph{without} higher-order terms.  For a free energy with the form of Eq.~(\ref{oseenfrank}) or~(\ref{newfreeenergy}), without higher-order terms, the conditions of Eq.~(\ref{necessary}) are necessary.

Summarizing the results of this section, we have derived a set of necessary inequalities~(\ref{necessary}), which are weaker than the Ericksen inequalities~(\ref{ericksen}).  If the necessary inequalities are violated, then the Ericksen inequalities are \emph{severely} violated, and the free energy may go to negative infinity.  This severe violation is forbidden (unless there are higher-order terms to stabilize the free energy).  However, there are intermediate regimes in which the Ericksen inequalities are violated but the necessary inequalities are satisfied.  In the following section, we consider what happens in those intermediate regimes.

\section{Intermediate regimes}

\begin{figure}
    \centering
    (a)\includegraphics[width=0.45\columnwidth]{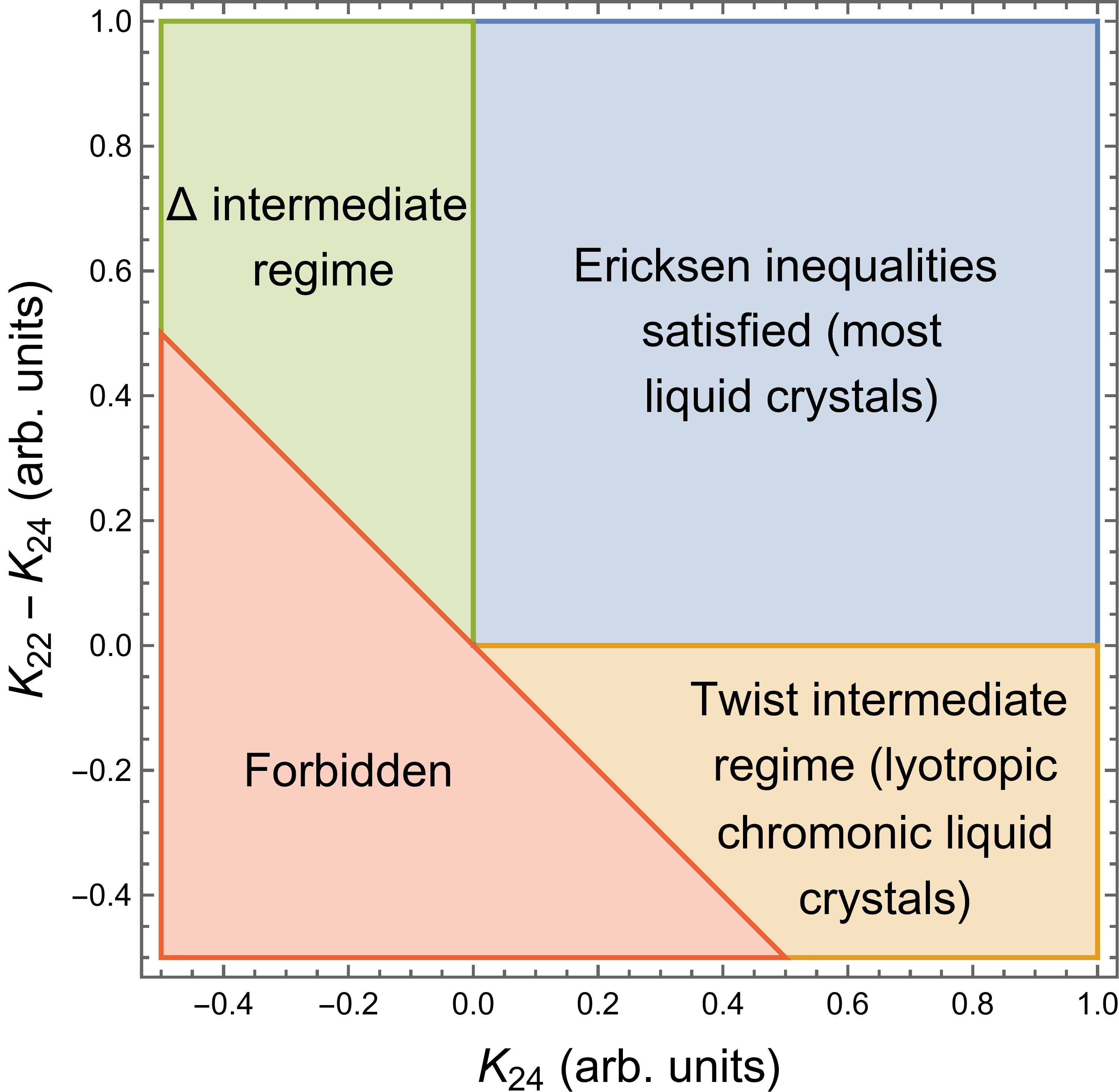}
    (b)\includegraphics[width=0.45\columnwidth]{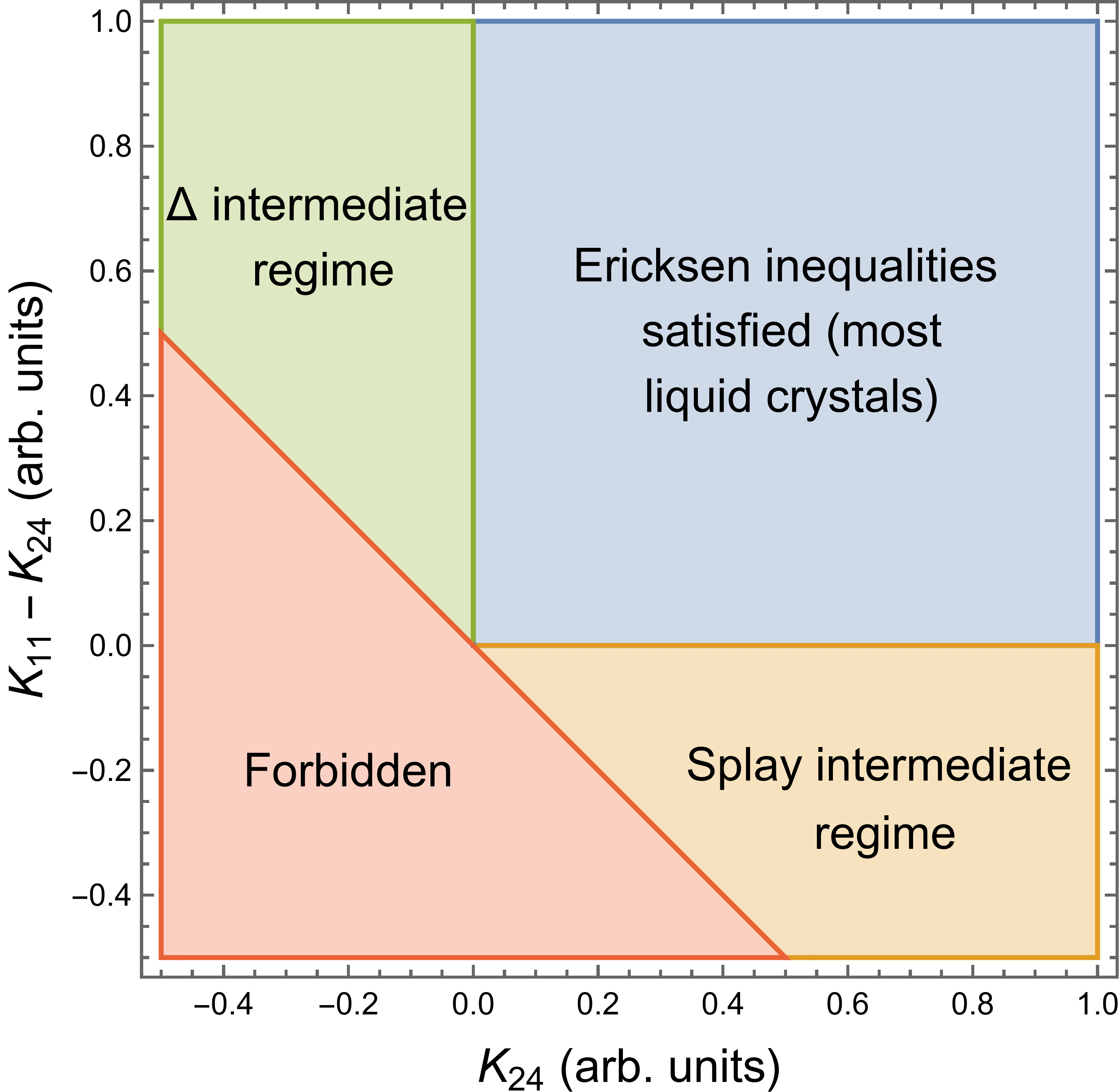}
    (c)\includegraphics[width=0.45\columnwidth]{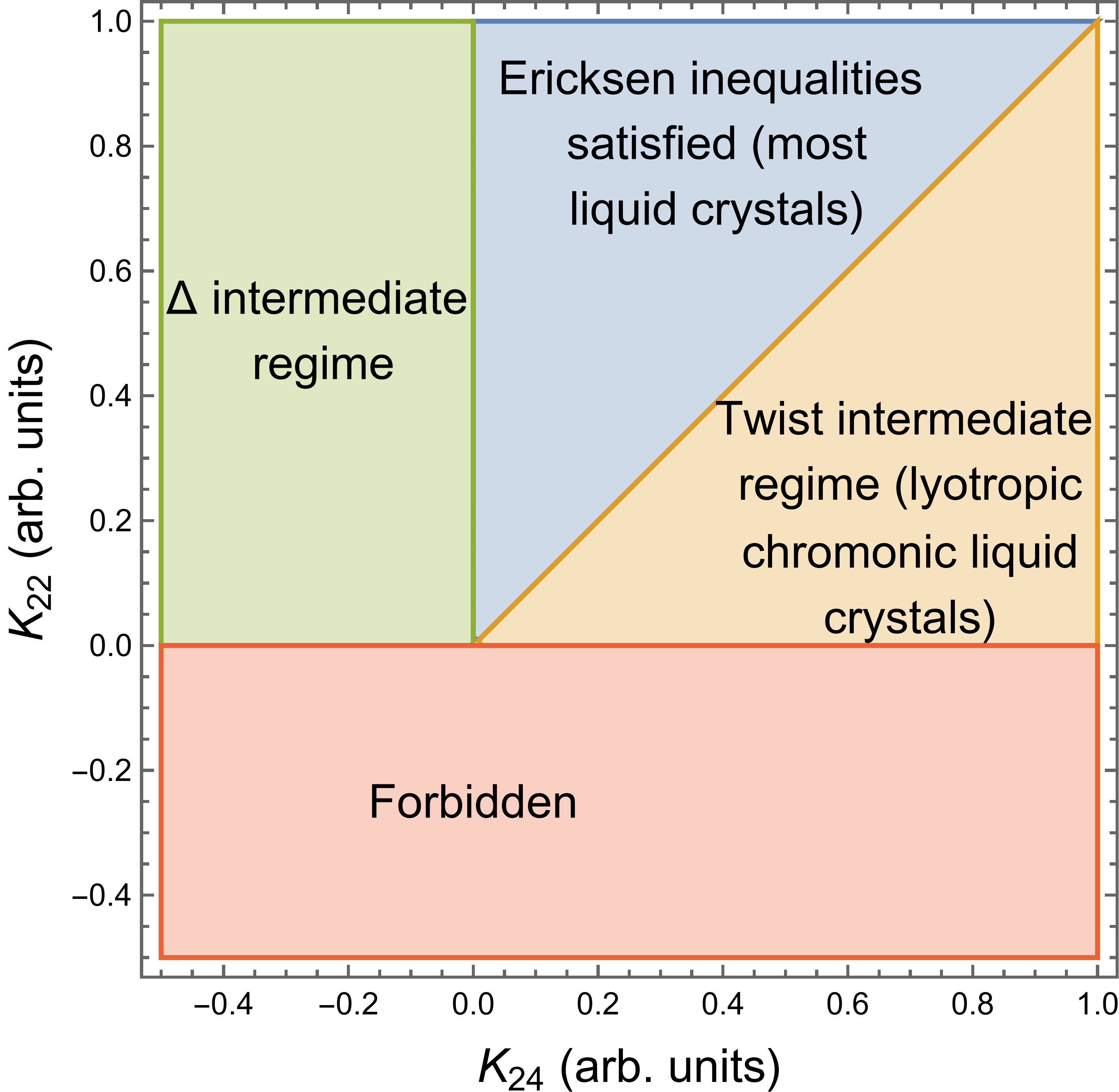}
    (d)\includegraphics[width=0.45\columnwidth]{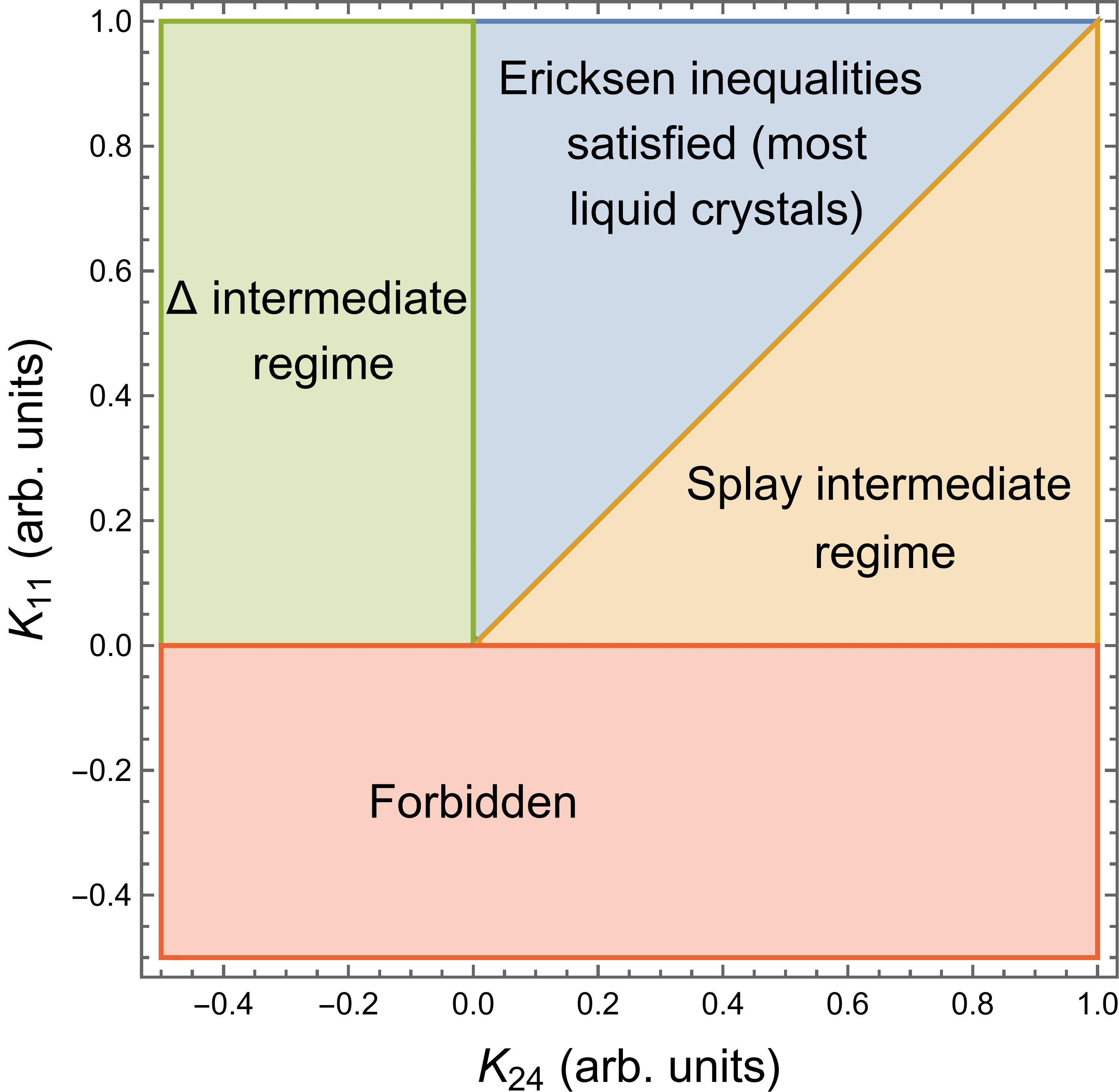}
    \caption{Possible regimes of the elastic constants.  (a,b) Expressed in terms of the fundamental elastic constants $(K_{11}-K_{24})$, $(K_{22}-K_{24})$, and $K_{24}$.  (c,d) Expressed in terms of the conventional elastic constants $K_{11}$, $K_{22}$, and $K_{24}$.  In parts (a,c), we vary $K_{22}$ and $K_{24}$, assuming that the splay and bend Ericksen inequalities are satisfied ($K_{11}>K_{24}$ and $K_{33}>0$).  In parts (b,d), we vary $K_{11}$ and $K_{24}$, assuming that the twist and bend Ericksen inequalities are satisfied ($K_{22}>K_{24}$ and $K_{33}>0$).}
    \label{fig:diagrams}
\end{figure}

Figure~1 shows schematic diagrams of possible regimes for the elastic constants.  The first row is expressed in terms of the fundamental elastic constants ${(K_{11}-K_{24})}$, $(K_{22}-K_{24})$, and $K_{24}$.  The second row provides the same information expressed in terms of the conventional elastic constants $K_{11}$, $K_{22}$, and $K_{24}$.  In each case, we can see that there is a blue regime where the Ericksen and the necessary inequalities are both satisfied, and a red regime where the Ericksen and necessary inequalities are both violated.  Between these regimes, there are intermediate regimes where the Ericksen inequalities are violated but the necessary inequalities are satisfied.  In particular, there are three intermediate regimes, which we label as:
\begin{itemize}
\item Twist intermediate regime:  $0<K_{22}<K_{24}$
\item Splay intermediate regime:  $0<K_{11}<K_{24}$
\item $\bm{\Delta}$ intermediate regime:  $K_{24}<0$, $K_{11}>0$, $K_{22}>0$
\end{itemize}
In this section, we concentrate on the twist intermediate regime, because it occurs experimentally in lyotropic chromonic liquid crystals.  At the end of the section, we briefly discuss the other two intermediate regimes, which have not yet been reported experimentally to our knowledge.  (We do not consider regimes in which more than one of the Ericksen inequalities are violated simultaneously.)

\subsection{Twist intermediate regime}

Suppose we are in the twist intermediate regime, where $0<K_{22}<K_{24}$, as has been reported in lyotropic chromonic liquid crystals.  In the free energy of Eq.~(\ref{newfreeenergy}), the coefficient of twist $T^2$ is negative, while all the other coefficients are positive.  Hence, the liquid crystal has a natural tendency toward twist; a twisted state should have a lower free energy than the uniform state.  However, the amount of twist is limited by the compatibility conditions.  If the director field has twist, it must have some of the other deformation modes, which have positive free energy.  Those other modes may stabilize the system, and prevent the free energy from going to negative infinity.

\begin{figure}
    \centering
    \includegraphics[width=0.45\columnwidth]{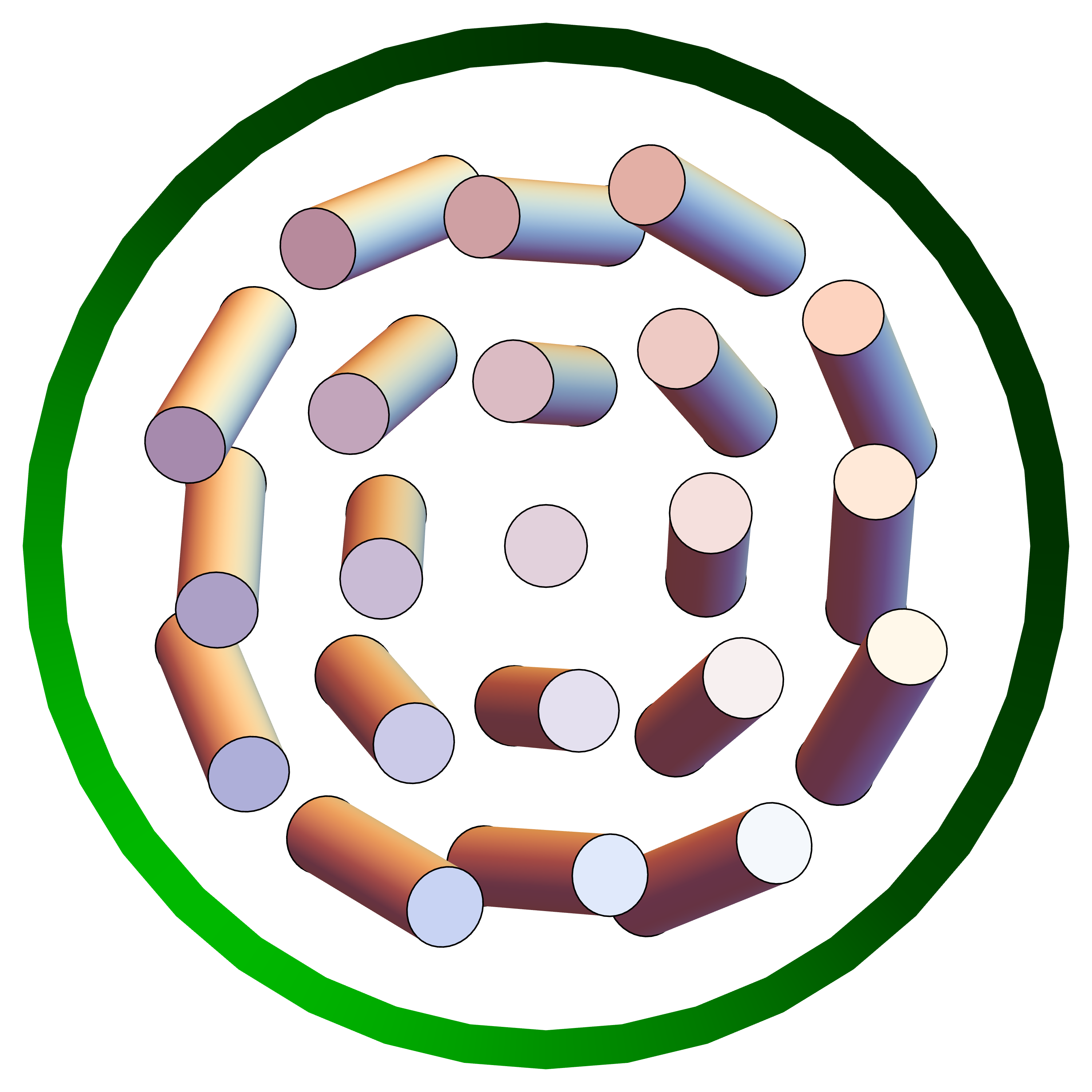}
    \caption{Twisted director field for a liquid crystal in a cylindrical capillary.  This structure is favored in the twist intermediate regime $0<K_{22}<K_{24}$.}
    \label{fig:twiststructure}
\end{figure}

To see these free energy considerations, suppose that the liquid crystal is in a cylindrical capillary with radius $R_\text{max}$, and suppose the boundary conditions on the surface are totally free.  Suppose the director field has the twisted structure shown in Fig.~\ref{fig:twiststructure}.  As a simple ansatz, it can be described in cylindrical coordinates $(\rho,\phi,z)$ by the equation
\begin{equation}
    \hat{\bm{n}}=\hat{\bm{z}}\cos\theta(\rho)+\hat{\bm{\phi}}\sin\theta(\rho),
    \label{ntwisted}
\end{equation}
with $\theta(\rho)=\alpha\rho$, for small $\alpha$.  This director field has twist of order $\alpha$, bend of order $\alpha^2 \rho$, $\bm{\Delta}$ mode of order $\alpha^3 \rho^2$, and zero splay.  If we put this ansatz into the free energy density and average over the cylindrical geometry, we obtain
\begin{equation}
    F_\text{average}=2(K_{22}-K_{24})\alpha^2
    +\frac{1}{12}\left[3K_{33}-8(K_{22}-K_{24})\right]R_\text{max}^2 \alpha^4
    +O(\alpha^6).
\end{equation}
This free energy has the usual form of a series expansion in powers of the twist order parameter $\alpha$.  When $(K_{22}-K_{24})>0$, the coefficient of the quadratic term is positive, and hence the minimum occurs at $\alpha=0$, which corresponds to an untwisted state.  When $(K_{22}-K_{24})=0$ there is a critical point, where the untwisted state has a symmetry-breaking transition to right- or left-handed twist.  Just below the critical point, the order parameter and free energy density scale as
\begin{equation}
    \alpha=\pm\frac{2}{R_\text{max}}\left[\frac{K_{24}-K_{22}}{K_{33}}\right]^{1/2},\quad
    F_\text{average}=-\frac{4(K_{24}-K_{22})^2}{K_{33}R_\text{max}^2}.
    \label{twistpredictions}
\end{equation}
The twist order parameter does not diverge, and the free energy does not go to negative infinity.  Rather, the system is stabilized by the compatibility requirement:  In order to have the favorable twist, the director field must also have some unfavorable bend (and a smaller amount of unfavorable $\bm{\Delta}$ mode).  These unfavorable deformation modes lead to a well-defined ground state, which has a free energy lower than the uniform state.

\begin{figure}
    \centering
    \includegraphics[width=0.6\columnwidth]{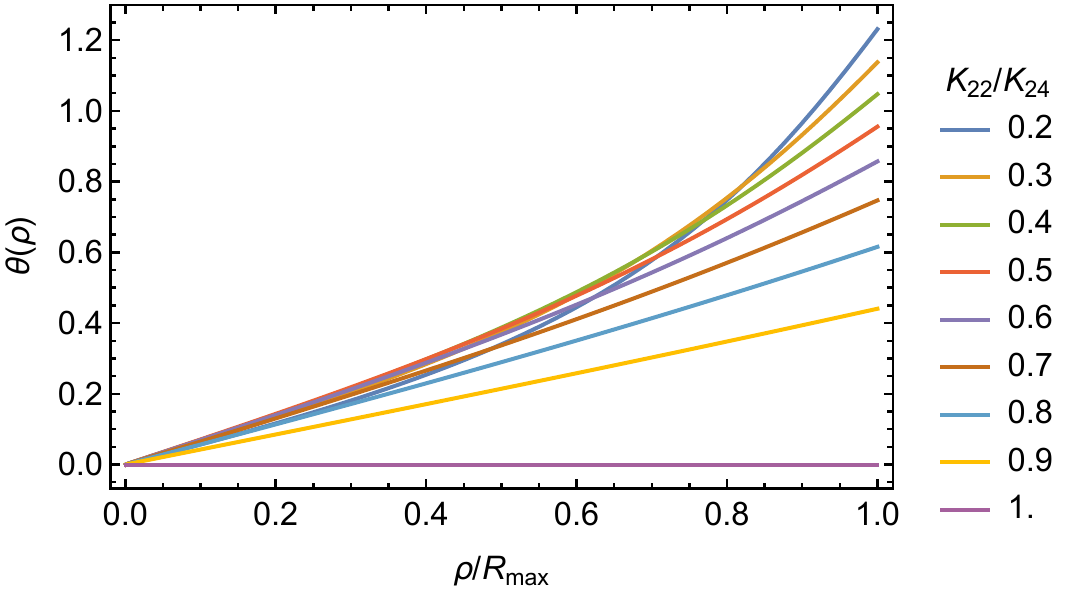}
    \caption{Numerical calculations of the director orientation $\theta(\rho)$ for a liquid crystal in a cylindrical capillary of radius $R_\text{max}$.  The ratio $K_{33}/K_{24}=2$ is fixed, and the ratio $K_{22}/K_{24}$ varies as indicated in the legend.}
    \label{fig:twistplots}
\end{figure}

As a check on this simple variational calculation, we consider a director field with the form of Eq.~(\ref{ntwisted}), but with an arbitrary function $\theta(\rho)$.  To minimize the free energy, we solve the Euler-Lagrange equation
\begin{equation}
    4K_{22}\rho^2 \theta''(\rho) + 4K_{22}\rho\theta'(\rho) - K_{22}\sin4\theta(\rho)
    = 8K_{33}\sin^3\theta(\rho)\cos\theta(\rho),
\end{equation}
with $\theta(0)=0$ and the free boundary condition
\begin{equation}
    2K_{22}R_\text{max}\theta'(R_\text{max})=(2K_{24}-K_{22})\sin2\theta(R_\text{max}).
\end{equation}
Note that $K_{24}$ does not enter the Euler-Lagrange equation, but it does enter the boundary condition.  We find numerical solutions with varying ratio $K_{22}/K_{24}$ from 1 down to 0.2, for fixed $K_{33}/K_{24}=2$.  The results are plotted in Fig.~\ref{fig:twistplots}.  When $K_{22}$ is just below $K_{24}$, the function is approximately linear, with the slope $\alpha$ as predicted in Eq.~(\ref{twistpredictions}).  When $K_{22}$ is substantially less than $K_{24}$, the shape of the function deviates from linearity, but still it has the same general form.  We can see that there is a well-defined ground state, with a twist that does not diverge, even deep in the twist intermediate regime.

From Eq.~(\ref{twistpredictions}), we see that the free energy depends on the system radius $R_\text{max}$ in a surprising way.  The average free energy density of the twisted state scales as $-1/R_\text{max}^2$, and it approaches the free energy density of the uniform state $F=0$ in the limit of $R_\text{max}\to\infty$.  Hence, the free energy per volume increases as the system size increases in the $(x,y)$ plane.  (By contrast, the free energy per volume is constant as the system size increases in the $z$ direction.)

This dependence on system size is quite unusual in liquid-crystal physics.  In typical liquid crystals, if we neglect surface effects, the free energy is extensive, meaning that the free energy per volume is constant as the system size increases.  In typical liquid crystals, if we include surface tension and surface-induced director distortions, the free energy per volume decreases as the system size increases; this behavior can be called sub-extensive.  By contrast, we now see that the free energy in the twist intermediate regime is super-extensive.  Meiri and Efrati~\cite{Meiri2021} have recently argued that super-extensive scaling of the free energy is a general characteristic of geometrically frustrated systems.  Our results for lyotropic chromonic liquid crystals provide an example of that phenomenon.

Whenever a system has super-extensive scaling of the free energy, one might ask whether it can reduce its total free energy by breaking into many smaller systems.  Specifically, for a lyotropic chromonic liquid crystal in a cylindrical capillary, one might ask whether it can break into a lattice of parallel tubes, with twist from the center to the edge of each tube.  In principle, this breakup would provide a large negative free energy for each tube, and hence reduce the total free energy.  However, the problem is how to connect the director field between neighboring tubes.  If we just put the tubes next to each other, and each tube has the structure shown in Fig.~\ref{fig:twiststructure}, then there would be discontinuities in the director field from tube to tube, and these discontinuities would cost a prohibitive amount of free energy.

\begin{figure}
    \centering
    (a)\includegraphics[width=0.45\columnwidth]{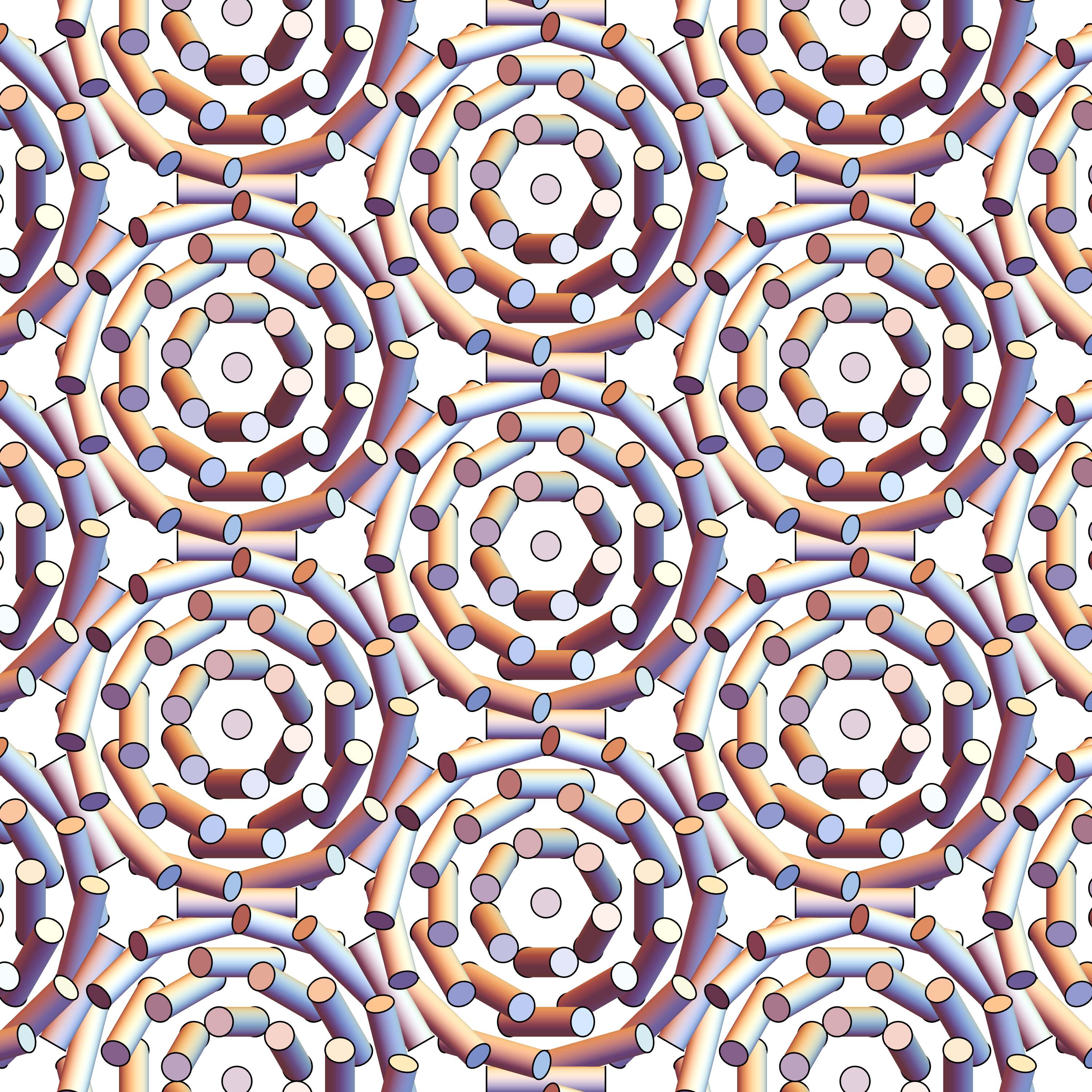}
    (b)\includegraphics[width=0.45\columnwidth]{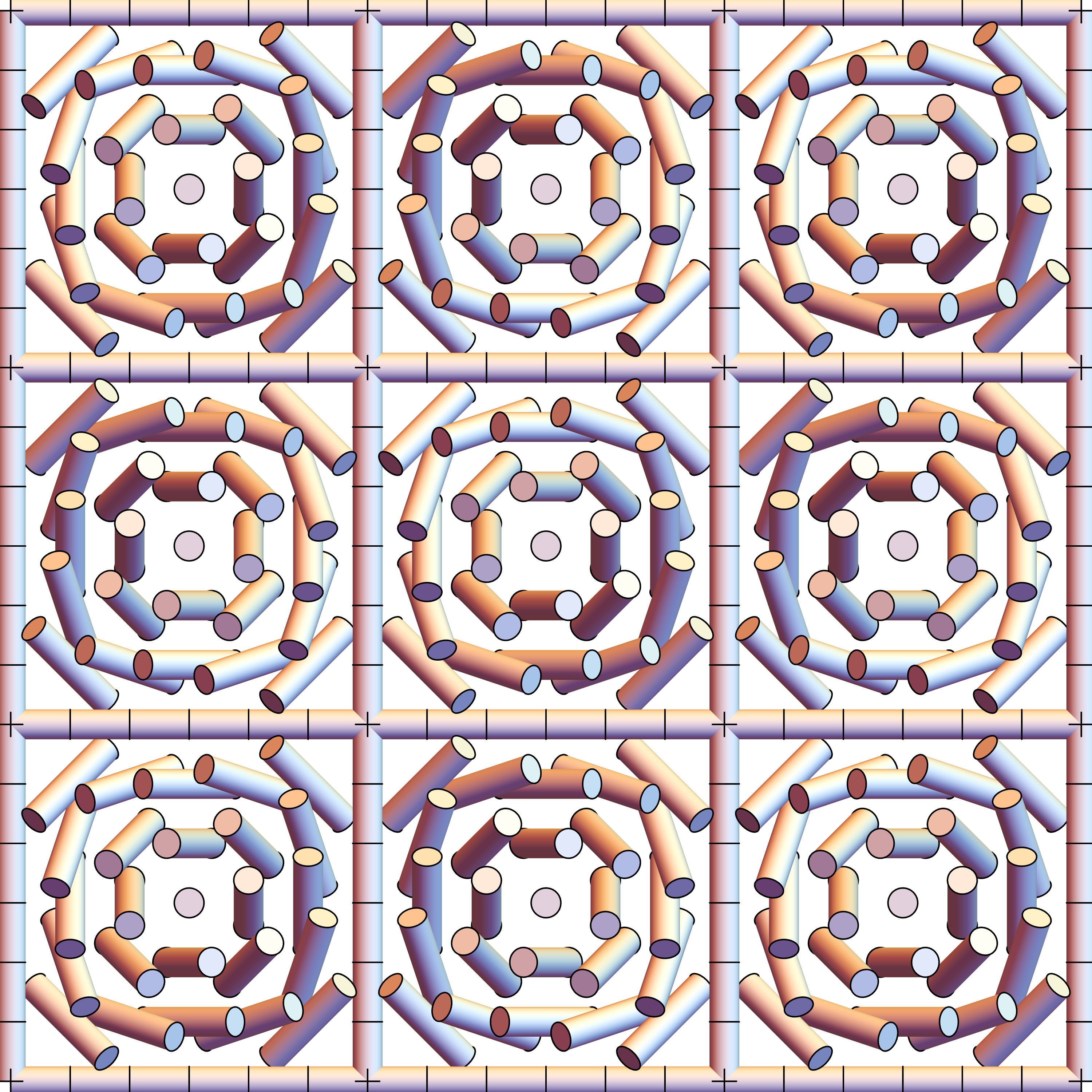}
    \caption{Possible structures for the director field in a liquid crystal where the disclination energy is low compared with $(K_{24}-K_{22})$, so that the free-energy benefit of the twist regions exceeds the free-energy cost of the disclinations between those regions.  (a)~Hexagonal lattice of twist tubes with uniform handedness.  (b)~Square lattice of twist tubes with alternating handedness.}
    \label{fig:lattices}
\end{figure}

Figure~\ref{fig:lattices} shows two possible structures for the director field in a lattice of parallel tubes.  Both of these structures can be regarded as arrays of half-skyrmions or merons, separated by disclinations.  They are essentially two-dimensional versions of blue phases.  In Fig.~\ref{fig:lattices}(a), all of the tubes have the same handedness and form a hexagonal lattice.  In Fig.~\ref{fig:lattices}(b), the tubes have alternating handedness and are arranged in a square lattice.  Half-skyrmion structures like Fig.~\ref{fig:lattices}(a) have been studied in the context of chiral liquid crystals, where they are stabilized by the favored twist arising from chirality~\cite{Nych2017,Duzgun2018}.  Here, we suggest that they might also form in achiral liquid crystals in the twist intermediate regime.  In that case, the free energy would become extensive, proportional to the number of tubes in the system.

For a chiral liquid crystal in the twist intermediate regime, the lattice of half-skyrmions would only be stable if the disclination energy is low compared with $(K_{24}-K_{22})$, so that the free-energy benefit from the twist regions exceeds the free-energy cost of the disclinations.  We have not seen any experimental evidence that this phenomenon actually occurs in lyotropic chromonic liquid crystals.  Rather, these materials appear to be in the regime of higher disclination energy, so that any cylindrical capillary has one twist region from the center to the edge, as in Fig.~\ref{fig:twiststructure}, and the free energy is super-extensive.

\subsection{Other intermediate regimes}

So far, we have considered liquid crystals in the twist intermediate regime, because that regime corresponds to experiments on lyotropic chromonic liquid crystals.  Two other theoretical possibilities are the splay intermediate regime and the $\bm{\Delta}$ intermediate regime.  We do not know of any liquid-crystal materials in those regimes.  However, we would like to make theoretical predictions for the director configurations, in case such materials should be discovered in the future.

First, consider the splay intermediate regime $0<K_{22}<K_{24}$.  In the free energy of Eq.~(\ref{newfreeenergy}), the coefficient of splay $S^2$ is negative, while all other coefficients are positive.  Hence, the liquid crystal has a tendency toward splay; a splayed state has a lower free energy than the uniform state.  However, as with the twist case, the amount of splay is limited by compatibility conditions, which may stabilize the system.

\begin{figure}
    \centering
    (a)\includegraphics[width=0.45\columnwidth]{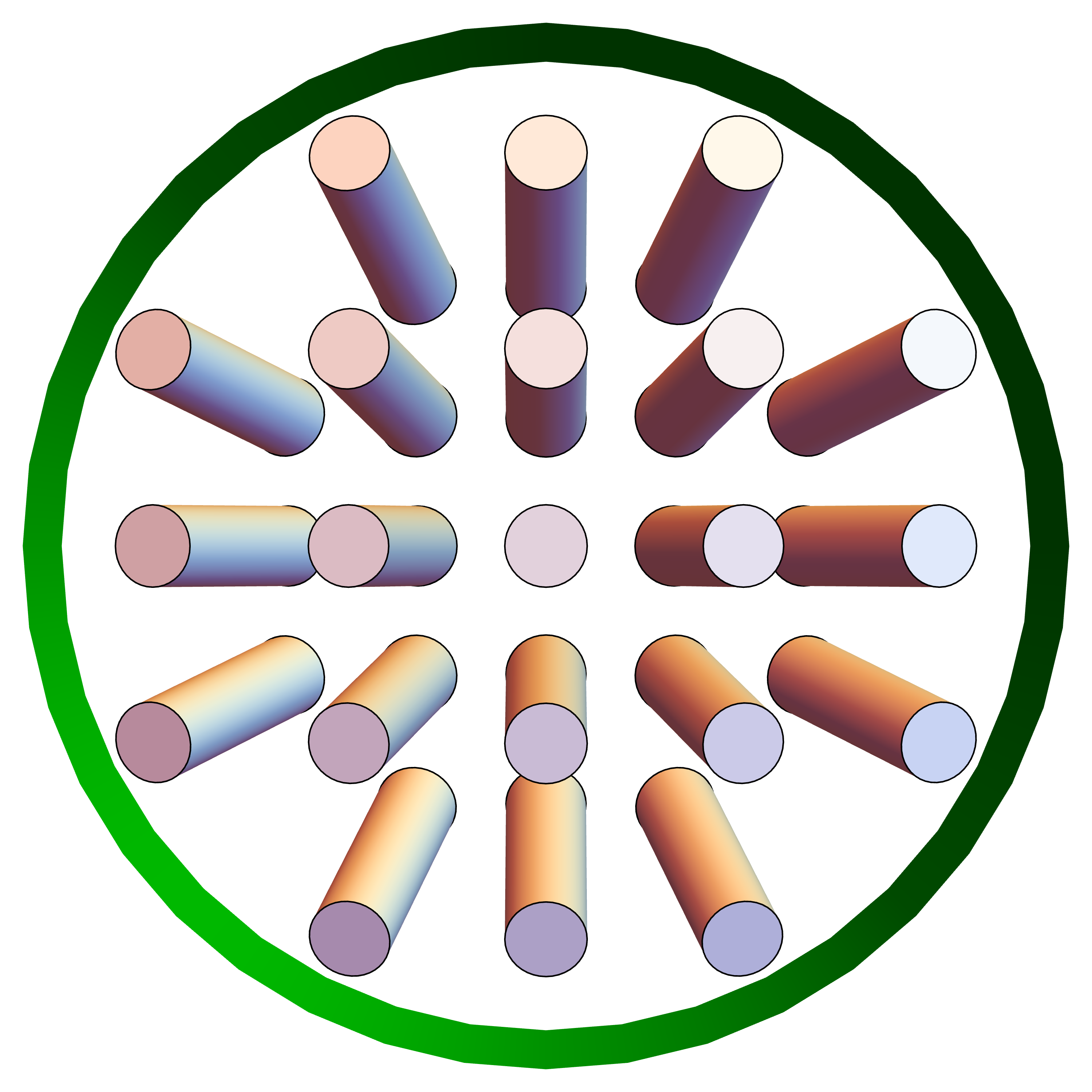}
    (b)\includegraphics[width=0.45\columnwidth]{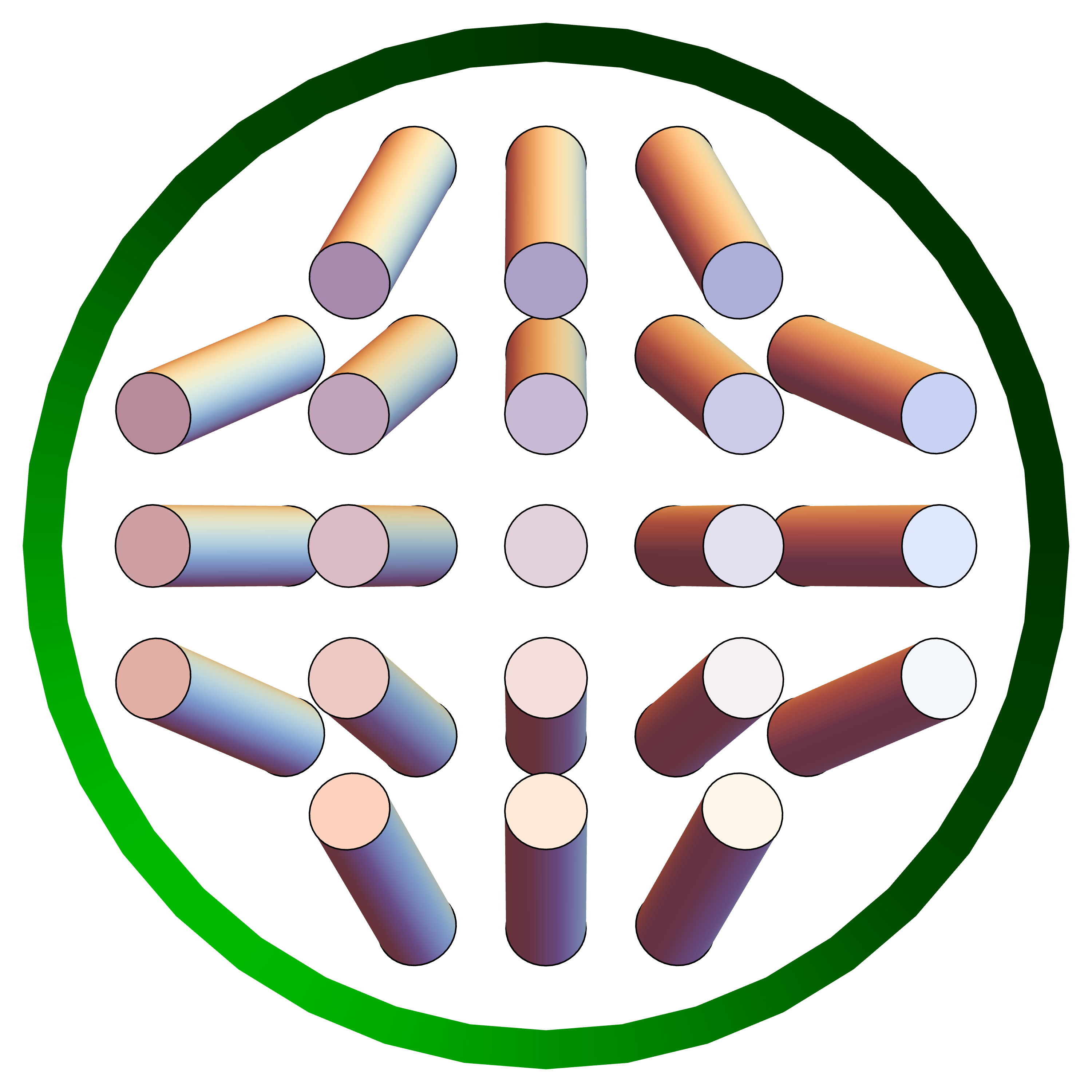}
    \caption{(a)~Director field in the splay intermediate regime $0<K_{11}<K_{24}$.
    (b)~Director field in the $\bm{\Delta}$ intermediate regime $K_{24}<0$.  (This example is drawn with the parameter $\phi_0 = 0$.)}
    \label{fig:splayandDeltastructures}
\end{figure}

As an example, suppose again that the liquid crystal is in a cylindrical capillary with radius $R_\text{max}$, with free boundary conditions, and suppose the director field has the splayed structure of Fig.~\ref{fig:splayandDeltastructures}(a).  It can be represented in cylindrical coordinates by the ansatz
\begin{equation}
    \hat{\bm{n}}=\hat{\bm{z}}\cos\theta(\rho)+\hat{\bm{\rho}}\sin\theta(\rho),
\end{equation}
with $\theta(\rho)=\alpha\rho$, for small $\alpha$.  This director field has splay of order $\alpha$, bend of order $\alpha^2 \rho$, $\bm{\Delta}$ mode of order $\alpha^3 \rho^2$, and zero twist.  By putting this ansatz into the free energy density and averaging over position in the cylinder, we obtain
\begin{equation}
    F_\text{average}=2(K_{11}-K_{24})\alpha^2
    +\frac{1}{12}\left[3K_{33}-8(K_{11}-K_{24})\right]R_\text{max}^2 \alpha^4
    +O(\alpha^6).
\end{equation}
This free energy is a series expansion in powers of the splay order parameter $\alpha$.  When $(K_{11}-K_{24})>0$, the quadratic coefficient is positive, and hence the minimum occurs at $\alpha=0$, which is the uniform state.  When $(K_{11}-K_{24})=0$, there is a critical point, where the uniform state has a symmetry-breaking transition to inward or outward splay, or equivalently, to splay up or down along the cylindrical axis.  Just below the critical point, the order parameter and free energy density scale as
\begin{equation}
    \alpha=\pm\frac{2}{R_\text{max}}\left[\frac{K_{24}-K_{11}}{K_{33}}\right]^{1/2},\quad
    F_\text{average}=-\frac{4(K_{24}-K_{11})^2}{K_{33}R_\text{max}^2}.
\end{equation}
All of these predictions for splay are exactly analogous to the predictions for twist in the previous section.

The same considerations apply in the $\bm{\Delta}$ intermediate regime $K_{24}<0$.  In the free energy of Eq.~(\ref{newfreeenergy}), the coefficient of $\Tr(\bm{\Delta}^2)$ is negative, and all other coefficients are positive.  In this case, the liquid crystal has a tendency toward the $\bm{\Delta}$ mode, so that a state with $\bm{\Delta}\not=0$ has a lower free energy than the uniform state.  Because the favorable $\bm{\Delta}$ deformation must be accompanied by other deformations, which are unfavorable, the system may still be thermodynamically stable.

To model the behavior in this regime, we again consider a liquid crystal in a cylindrical capillary with free boundary conditions, and now assume the director field
\begin{equation}
    \hat{\bm{n}}=\hat{\bm{z}}\cos\theta(\rho)+\hat{\bm{\rho}}\sin\theta(\rho)\cos2(\phi-\phi_0)
    -\hat{\bm{\phi}}\sin\theta(\rho)\sin2(\phi-\phi_0),
\end{equation}
with $\theta(\rho)=\alpha\rho$, for small $\alpha$.  This structure is illustrated in Fig.~\ref{fig:splayandDeltastructures}(b); note that the outward tilt is at orientations of $2\phi_0$ and $2\phi_0+\pi$ with respect to the $x$-axis.  It has $\bm{\Delta}$ mode of order $\alpha$, bend of order $\alpha^2 \rho$, and splay and twist of order $\alpha^3 \rho^2$.  We put this ansatz into the free energy density, and average over position, to obtain
\begin{equation}
    F_\text{average}=2K_{24}\alpha^2
    +\frac{1}{12}\left(3K_{33}-8K_{24}\right)R_\text{max}^2 \alpha^4
    +O(\alpha^6).
\end{equation}
This free energy is a power series in $\alpha$, which can now be regarded as an order parameter for the $\bm{\Delta}$ deformation.  When $K_{24}>0$, the minimum occurs at the uniform state $\alpha=0$.  When $K_{24}=0$, there is a critical point, where the uniform state has a symmetry-breaking transition to some non-zero $\bm{\Delta}$ deformation, with an arbitrary orientation $\phi_0$.  Just below the critical point, the order parameter and free energy density scale as
\begin{equation}
    \alpha=\pm\frac{2}{R_\text{max}}\left[\frac{-K_{24}}{K_{33}}\right]^{1/2},\quad
    F_\text{average}=-\frac{4K_{24}^2}{K_{33}R_\text{max}^2}.
\end{equation}
These predictions are quite analogous to the twist and splay cases.

All of our discussion about system-size dependence and geometric frustration in the twist intermediate regime applies also to the splay and $\bm{\Delta}$ intermediate regimes.  In particular, if the disclination energy is very low, the liquid crystal might break up into domains of the favored mode separated by disclinations.  For the splay case, we expect that the domain structure might resemble Fig.~\ref{fig:lattices}(a) or~\ref{fig:lattices}(b), but with the entire director field rotated by $\pi/2$ about the $z$-axis, so that twist is transformed into splay.  For the $\bm{\Delta}$ case, we have not yet investigated the possible domain structures.

\section{Discussion}

In this article, we have identified three distinct regimes of elastic constants in nematic liquid crystals, which are indicated schematically in Fig.~\ref{fig:diagrams}.  In the regime where the Ericksen inequalities~(\ref{ericksen}) are satisfied, all four director deformation modes cost some positive free energy.  For that reason, the ground state of a bulk liquid crystal has a uniform director field.  By contrast, in the forbidden regime where the necessary inequalities~(\ref{necessary}) are violated, some physically realizable director field has a negative free energy, and hence the free energy~(\ref{oseenfrank}) or~(\ref{newfreeenergy}) is thermodynamically unstable.  It can only be stabilized by extra terms with higher powers of the director gradients, or higher-order derivatives of the director field, or couplings with other order parameters.  In this forbidden regime, the liquid crystal may form a modulated structure with no singularities in nematic order, such as a twist-bend nematic phase or a splay nematic phase, as studied in much recent research~\cite{Dozov2001,Shamid2013,Rosseto2020,Selinger2022,Rosseto2022}.

Between the Ericksen regime and the forbidden regime, there is an intermediate regime where the Ericksen inequalities are violated but the necessary inequalities are satisfied.  In this intermediate regime, one of the director deformation modes---twist, splay, or $\bm{\Delta}$---has a negative free energy.  However, the total free energy is stabilized by geometric compatibility constraints, which require that any physically realizable director field must have a combination of the favored mode with other, unfavorable modes.  The intermediate regime has surprising properties.  In a finite cylindrical geometry, the liquid crystal minimizes its free energy by forming a nonuniform director field, as shown in Fig.~\ref{fig:twiststructure}, \ref{fig:splayandDeltastructures}(a), or \ref{fig:splayandDeltastructures}(b).  As the cylinder radius increases, the total free energy increases super-extensively.  The super-extensive growth can only be avoided if the system adds disclinations, as in Fig.~\ref{fig:lattices}, provided that the disclination energy is low enough.

Remarkably, the twist intermediate regime actually occurs in lyotropic chromonic liquid crystals, according to experimental measurements of the elastic constants~\cite{Nayani2015,Davidson2015,Fu2017,Javadi2018}.  Hence, the behavior discussed here is not just a theoretical speculation, but can be studied in the laboratory.  In particular, we emphasize three consequences of this theory for experiments on lyotropic chromonic liquid crystals.

First, these materials should be sensitive to the geometry of their container.  In this article, we have seen the effects of changing the radius $R_\text{max}$ of a cylindrical capillary.  We would expect equally important effects of changing the shape of the container.  The favorable twist deformation naturally fills up a region that is roughly circular in the plane perpendicular to the director.  If the cross section of the cell is not circular, then the director field must adapt in a complex way, perhaps by forming circular domains of twist separated by untwisted regions.  As an example, the experiments of Ref.~\cite{Fu2017} put lyotropic chromonic liquid crystals into capillaries of rectangular cross section and find complex director configurations.

Second, these materials have an unusual relationship between the free energies of double twist, cholesteric single twist, and a uniform state.  In typical chiral liquid crystals, double twist is preferred over single twist, and single twist is preferred over a uniform state.  By contrast, in lyotropic chromonic liquid crystals, double twist is preferred over a uniform state, and a uniform state is preferred over single twist.  Hence, the materials should particularly avoid cholesteric single twist.  Indeed, the structures in Ref.~\cite{Fu2017} include double-twist regions and monodomain uniform regions.

Third, we anticipate that these materials should be particularly compatible with impurities, such as dust or other colloidal particles.  Any particles will break up the liquid crystals into smaller volumes, and will allow the director field to break into more regions of double twist.  Effectively, the particles could play the same role as the disclinations in the structures of Fig.~\ref{fig:lattices}.  Hence, the negative free energy of the twist domains could offset the positive free energy of contact between the particles and the liquid crystal.  Distortions induced by such particles could make these materials difficult to align.  A recent study suspends rod-like particles in a lyotropic chromonic liquid crystal, and finds an anomalous twisted alignment of the rods with respect to the director field~\cite{Ettinger2022}.  We speculate that the spontaneous twist discussed here may be involved in that experiment.

\backmatter

\bmhead{Acknowledgments}

This work was supported by National Science Foundation Grant DMR-1409658.

\bibliography{sn-bibliography}

\end{document}